\documentstyle[sprocl]{article}

\def\Journal#1#2#3#4{{#1} {\bf #2}, #3 (#4)}

\def\NPB{{\em Nucl. Phys.} B}
\def\PLB{{\em Phys. Lett.}  B}

\def\al{\alpha}

\def\be{\begin{equation}}
\def\ee{\end{equation}}
\def\bea{\begin{eqnarray}}
\def\eea{\end{eqnarray}}

\begin{document}

\title{LIGHT VECTOR MESON PHOTOPRODUCTION AT HIGH $|t|$}

\author{D. YU. IVANOV}

\address{Institute of Mathematics,\\ 630090 Novosibirsk, Russia\\E-mail: d-ivanov@math.nsc.ru}

\maketitle\abstracts{ We have studied in perturbative QCD all independent 
helicity amplitudes describing
the photoproduction of light vector mesons at large $t$.
We found a new hard production mechanism 
which is related to the possibility for a real photon 
to fluctuate into a massless $q\bar q$ pair in a chiral-odd 
spin configuration. 
Each helicity amplitude is given as a sum of   
chiral-even contribution (when the helicities
of quark and antiquark are antiparallel) and this additional
chiral-odd part (where 
the helicities of quark and antiquark are parallel). 
The chiral-odd contribution is large, it leads to a dominance of 
the non spin-flip amplitude in a very broad region of intermediately 
high $|t|$.}

\section{Introduction}

My talk is based on our recent publication \cite{pub}. 
We derived pQCD predictions for helicity 
amplitudes  of the high $|t|=-{\bf q}^2$ diffractive light 
vector meson photoproduction process. 
Initial proton disintegates into hadron system $X$ which 
is separated from the produced vector meson $V$ by the rapidity gap $\eta_0$.  
The cross section for this reaction  
can be related to those for the photoproduction of $V$
off  a
quark and a gluon via the gluon and quark densities in a proton
$G(x,t)$ and $q(x,t)$:
\bea
\label{def2}
\frac{d^2\sigma (\gamma p\to VX)}{dtdx}&=&
\sum_f\left(q(x,t)+\bar q(x,t)\right)
\frac{d\sigma(\gamma q\to Vq)}{dt}+\nonumber\\
&&G(x,t)\frac{d\sigma(\gamma G\to VG)}{dt};\quad
x=\frac{4 |t|}{s}\cosh^2\frac{\eta_0}{2}.\label{strf}
\eea

There are three independent helicity amplitudes for the parton 
subprocess $\gamma q\to Vq$,
the first and the second indices are the helicities 
of photon and vector meson respectively
$$
M_{+\;+} (M_{-\;-}=M_{+\;+})\ , \  
M_{+\;0} (M_{-\;0}=-M_{+\;0}) \ , \
M_{+\;-} (M_{-\;+}=M_{+\;-}) \ .
$$

Within usual  perturbation theory a photon 
can split only into a chiral-even
(the helicities of the quark and the
antiquark are antiparallel)
massless quark pair.
The violation of chiral symmetry, which is well known to be a 
soft QCD
phenomenon, generates a nonperturbative 
chiral-odd component of the 
real photon wave function.
The interaction of this additional chiral contribution can however
be described in  pQCD since high $t$ quark-dipole scattering chooses
a $q\bar q$ configuration with  small transverse 
interquark distances. As a result  the chiral-odd contribution
 can be factorized into
a convolution of two nonperturbative 
photon and vector meson light-cone wave functions with the hard scattering 
amplitude.
The chiral-odd wave function of a real photon have a
similar form as the chiral-odd wave function of a vector meson. 
The photon dimentional coupling constant $f_\gamma$ (wich is similar
to vector meson chiral-odd constant $f_V^T$) is
a product of the quark condensate
$<{\bar q}q>$ and its magnetic susceptibility \cite{suscep}, 
\begin{equation}
f_\gamma = <{\bar q}q>\,\chi \approx 70 MeV\;.
\label{fgamma}
\end{equation}
Both parameters   $<{\bar q} q> $ and $\chi$ describing QCD vacuum
have been tested in various QCD sum rule applications.

\section{Results and Discussion}

The helicity amplitudes of the process
$\gamma q\to V q$ are the sums of  the 
chiral-even and the chiral-odd
contributions $$
M_{\lambda_1\;\lambda_2} = M^{even}_{\lambda_1\;\lambda_2} + M^{odd}
_{\lambda_1\;\lambda_2} \ .
$$
At asymptotically high $|t|$ the dominant 
helicity amplitude is $M_{+\;0}$. Its chiral-even 
part $M^{even}_{+\;0}$  
has the minimal, $\sim 1/{\bf q}^3$, suppression.
\begin{equation}
M^{even}_{+\,0} = -is\al_s^2\,\frac{32\pi}{3\sqrt{2}}\,eQ_V f_V /q^3;\ .
\label{T0even3}
\end{equation} 
We present our results for helicity amplitudes as the ratios
$M_{\lambda_1\;\lambda_2}/M^{even}_{+\;0}$   
\begin{equation}
\frac{M_{+\;0}}{M^{even}_{+\;0}}= 1- \frac{24\pi^2\,f_\gamma f_V^T
m_V}{3 f_V\,{\bf q}^2} \left( \ln \frac{1 - u_{min}}{u_{min}} 
- 2(1-2u_{min})\right)
\label{d0}
\end{equation}
\begin{equation}
\frac{M_{+\;+}}{M^{even}_{+\;0}}= 
\frac{m_V}{|{\bf q}|\sqrt{2}}\left(\ln \frac{1 -u_{min}}{u_{min}} -1 +
2u_{min}  \right) + \frac{24\pi^2f_\gamma f_V^T}{3 f_V
|{\bf q}|\sqrt{2}}
\label{d+}
\end{equation}
\begin{equation}
\frac{M_{+\;-}}{M^{even}_{+\;0}}= \frac{3\,m_V}{\sqrt{2}|{\bf q}|} -
\frac{48\pi^2f_\gamma f_V^T m_V^2}{3 f_V |{\bf q}|^3 \sqrt{2}} 
\left(2 \ln \frac{1 -u_{min}}{u_{min}} -3(1 -2u_{min})     \right)
\label{d-}
\end{equation}

The chiral-odd contributions to the amplitudes 
are accompanied with large numerical coefficients.
In the case of $M_{+\;+}$ the chiral-even and the chiral-odd parts
of (\ref{d+}) add  with the same signs. In  contrast,
the chiral-even and the chiral-odd parts 
of $M_{+\;0}$ or $M_{+\;-}$ 
enters with  opposite sign which leads to an effective 
reduction of these amplitudes for  intermediate $|t|$. 
Though the  chiral-even and the chiral-odd contributions
to $M_{+\;+}$ are of the same order with respect to 
$1/{|{\bf q}}|$ counting the chiral-odd one can be dominant
up to very large $|t|$. According to (\ref{d+})
for the $t$ range $3 \; \div \;8\mbox{ GeV}^2$
 the chiral-even part constitutes only $10\;\div \;20\%$ of 
the $M_{+\;+}$ amplitude and 
for $|t| \approx 100 \mbox{GeV}^2\;$ $M^{even}_{+\;+}
\approx 0.72\,M^{odd}_{+\;+}$.
Due to a large compensation between chiral-even and 
chiral-odd parts of $M_{+\;0}$ and $M_{+\;-}$ 
the non spin-flip $M_{+\;+}$ amplitude dominates strongly
in the region of intermediately high $|t|$. 
$M_{+\;0}$ will exceed $M_{+\;+}$
only at $|t|> 40(\mbox{ GeV})^2$. 
For the $|t|$ interval $3\mbox{ GeV}^2 \; \div \; 8\mbox{ GeV}^2$
\begin{equation}
\frac{M_{+\;0}}{M_{+\;+}}\sim 0.25\; \div\; 0.35 \ , \
\frac{M_{+\;-}}{M_{+\;+}}\sim -\,0.02\; \div \; \,0.04
\label{df0}
\end{equation}

Both chiral-even and chiral--odd parts of the 
amplitudes were calculated in the leading order of $1/|{\bf q}|$
expansion.  As usual in the 
QCD approach to any exclusive reaction, the question about the region
of applicability of these results is open
untill the power corrections have not studied.
In our case the situation is more difficult because
the factorization of the amplitude into  hard and soft parts 
is violated
for the chiral-even part of $M_{+\;+}$ and the chiral-odd
parts of $M_{+\;0}$ and $M_{+\;-}$. In these cases 
the corresponding integrals over the quark longitudinal 
momentum $u$ contain the end point logarithmic singularities. 
At  present we simply restrict 
the corresponding $u$ integrals to the 
interval $[1- u_{min}, u_{min}], u_{min}=m_V^2/{\bf q}^2$,  
which corresponds to the contribution 
of the hard region only. 
The good news is, however,
that the chiral-odd part $M_{+\;+}^{odd}$ of the dominant 
in the intermediatly large $|t|$ region  helicity amplitude 
$M_{+\;+}$ is free from this end-point singularity.
This chiral-odd part is numerically considerably larger 
than the hard contribution to its chiral-even counterpart
$M_{+\;+}^{even}$. Therefore we can expect that 
the relative uncertainty related with the uncalculated soft contributions
to $M_{+\;+}^{even}$ is small.

\section*{Acknowledgments}
This work was supported by the Foundation
Universities of Russia (grant 015.02.01.16).

\end{document}